\documentclass{article}
\usepackage{graphicx}
\usepackage{amsmath}

\newtheorem{theorem}{Theorem}

\begin{document}

\begin{center}
{\LARGE Noether's Theorems and Gauge Symmetries}

\bigskip

\textit{Katherine Brading}

\textit{St. Hugh's College, Oxford, OX2 6LE}

katherine.brading@st-hughs.ox.ac.uk

and

\textit{Harvey R. Brown}

\textit{Sub-faculty of Philosophy, University of Oxford, }

\textit{10 Merton Street, Oxford OX1 4JJ}

harvey.brown@philosophy.ox.ac.uk

\bigskip

\textit{August 2000}

\bigskip
\end{center}

{\small Consideration of the Noether variational problem for any theory
whose action is invariant under global and/or local gauge transformations
leads to three distinct theorems. These include the familiar Noether
theorem, but also two equally important but much less well-known results. We
present, in a general form, all the main results relating to the Noether
variational problem for gauge theories, and we show the relationships
between them. These results hold for both Abelian and non-Abelian gauge
theories.}

\bigskip

\section{Introduction}

\bigskip

There is widespread confusion over the role of Noether's theorem in the case
of local gauge symmetries,\footnote{%
The confusion goes beyond gauge symmetries; for discussion in this journal
of this point see, for example, Munoz (1996). The results presented in this
paper extend straightforwardly to such cases.} as pointed out in this
journal by Karatas and Kowalski (1990), and Al-Kuwari and Taha (1991).%
\footnote{%
A key issue addressed in these papers is why no further conserved quantities
arise from local gauge symmetries than already arise from global gauge
symmetries. The reason for this is clearly seen in what follows.} In our
opinion, the main reason for the confusion is failure to appreciate that
Noether offered two theorems in her 1918 work. One theorem applies to
symmetries associated with finite dimensional Lie groups (global
symmetries), and the other to symmetries associated with infinite
dimensional Lie groups (local symmetries); the latter theorem has been
widely forgotten. Knowledge of Noether's `second theorem' helps to clarify
the significance of the results offered by Al-Kuwari and Taha for local
gauge symmetries, along with other important and related work such as that
of Bergmann (1949), Trautman (1962), Utiyama (1956, 1959), and Weyl (1918,
1928/9). In this paper we present all the key results concerning Noether's
theorems for global and local gauge symmetries - including those which go
beyond Noether's own derivations - in the form of three simple theorems and
their consequences. In the process, we highlight several important and
useful results that have been largely overlooked, and aim to help bring an
end to the confusion over this issue.

\bigskip

The results we present are all derivable from the variational problem stated
in section \ref{Basis}. Section 3 considers global gauge symmetry, and
states the associated (and familiar) Noether theorem. Sections 4-6 address
local gauge symmetry. In section 4 we state the second Noether theorem, and
give an example of its applications. Section 5 addresses the application of
the first Noether theorem to global subgroups of local gauge groups.
Finally, in section 6, we discuss the paper of Al-Kuwari and Taha (1991).
Their paper is based on results due to Utiyama (1956); we summarise these in
the form of a theorem, and highlight what we what believe to be the most
important aspect of Al-Kuwari and Taha's paper.

\bigskip

\section{Basis of the Noether Theorems\label{Basis}}

\bigskip

It is useful to compare Noether's variational \textit{problem} with the more
familiar Hamilton's \textit{principle}.

\bigskip

Consider a Lagrangian density $L$ depending on $N$ distinct fields $\psi _{i}
$ ($i=1,...,N$) and their first derivatives, written as $L=L(\psi
_{i},\partial _{\mu }\psi _{i},x^{\mu }).\footnote{%
The restriction of $L$ to $L=L(\psi _{i},\partial _{\mu }\psi _{i},x^{\mu })$
and no higher derivatives of $\psi _{i}$ is for convenience. The
generalisation of everything that follows to higher derivatives is
straightforward.}$ The action $S$ is defined as $S=\int Ld^{4}x$ over some
compact region of space-time. Hamilton's principle, defined for a particular
field, $\psi _{k}$, requires the action to be extremal (that is, $\delta S=0,
$ where $\delta S$ is the first order functional variation in $S)$ for
\textit{arbitrary} variations of $\psi _{k}$ which vanish on the boundary.
As is well-known, the necessary and sufficient condition for this principle
to hold is satisfaction of the Euler-Lagrange equations for $\psi _{k}.$
(Note that the principle may not apply to all the fields on which a given
Lagrangian depends. For an example, see section 6, footnote 14.)

\bigskip

Noether's variational problem (VP) can be posed as follows:

\bigskip

\textit{What general conditions must hold in order that a given variation of
the dependent and/or independent variables leaves the action invariant, and
hence }$\delta S=0,$\textit{\ where }$\delta S$\textit{\ may now contain a
boundary term?}

\bigskip

Clearly this variational problem is importantly different from Hamilton's
principle, both in the sets of variations considered, and in purpose.

\bigskip

The general solution of VP is the following condition:\footnote{%
Details of the derivation can be found in Noether (1918), Doughty (1990, p.
338) and Brading and Brown (2000), for example. Note that VP, and the
resulting expression, may be generalised to allow that the Lagrangian may
pick up a divergence term under the variation. This is needed for Galilean
boosts in particle mechanics, for example. For further discussion of this
point see Doughty (1990) and Brading and Brown (2000).}
\begin{equation}
\sum_{i}\left[ \Psi \right] _{i}\delta _{0}\psi _{i}\equiv -\sum_{i}\partial
_{\mu }B_{i}^{\mu }  \label{NoetherGenExp}
\end{equation}
where

\begin{enumerate}
\item  $\left[ \Psi \right] _{i}$ is the `Lagrange expression' associated
with the field $\psi _{i}$:
\begin{equation}
\left[ \Psi \right] _{i}:=\frac{\partial L}{\partial \psi _{i}}-\partial
_{\mu }\left( \frac{\partial L}{\partial \left( \partial _{\mu }\psi
_{i}\right) }\right)
\end{equation}
i.e., $\left[ \Psi \right] _{i}=0$ are the Euler-Lagrange equations for $%
\psi _{i}$;

\item  the variation of each $\psi _{i}$ (denoted by $\delta \psi _{i})$ is
composed of the direct variation in $\psi _{i}$ plus that which arises as a
consequence of the variation in $x^{\mu }$:
\begin{equation}
\delta \psi _{i}=\delta _{0}\psi _{i}+\left( \partial _{\mu }\psi
_{i}\right) \delta x^{\mu };
\end{equation}
and

\item  the form of $B_{i}^{\mu }$ is:
\begin{equation}
B_{i}^{\mu }:=\left( L\delta x^{\mu }+\frac{\partial L}{\partial \left(
\partial _{\mu }\psi _{i}\right) }\delta _{0}\psi _{i}\right) .
\end{equation}
\end{enumerate}

\bigskip

Throughout this paper, we use the symbol `$\equiv $' to indicate equations
that are derived without making use of any Euler-Lagrange equations, and the
Einstein convention to sum over repeated Greek indices, all other summations
being explicit.

\bigskip

In the case of gauge transformations in field theory we are concerned with
transformations of the fields only (i.e., the dependent variables), and not
transformations of the space-time coordinates (the independent variables),
and hence we ignore terms in $\delta x^{\mu }.$ In this case, we have $%
\delta _{0}\psi _{i}=\delta \psi _{i}$ and (\ref{NoetherGenExp}) becomes
\begin{equation}
\sum_{i}\left[ \Psi \right] _{i}\delta \psi _{i}\equiv -\sum_{i}\partial
_{\mu }C_{i}^{\mu }  \label{GenExpGauge}
\end{equation}
where
\begin{equation}
C_{i}^{\mu }:=\frac{\partial L}{\partial \left( \partial _{\mu }\psi
_{i}\right) }\delta \psi _{i}.  \label{B(fields only)}
\end{equation}

\bigskip

This is the first stage in the derivation of all three theorems presented in
this paper.\footnote{%
Generalisations of all these theorems, based on (\ref{NoetherGenExp}) rather
than (\ref{GenExpGauge}), are straightforward. See Brading and Brown, 2000.}

\bigskip

When the Euler-Lagrange field equations are satisfied for all of the fields
on which the Lagrangian depends, the `Lagrange expressions' $\left[ \Psi %
\right] _{i}$ vanish, so $\sum_{i}\left[ \Psi \right] _{i}=0$,\footnote{%
The assumption that the Euler-Lagrange field equations are satisfied yields $%
\sum_{i}\left[ \Psi \right] _{i}=0$ \textit{iff} all the fields on which the
Lagrangian depends satisfy Euler-Lagrange equations. The significance of
this remark will be made clear in section \ref{FieldEqns}, below.} and from (%
\ref{GenExpGauge}) we have the continuity equation
\begin{equation}
\sum_{i}\partial _{\mu }C_{i}^{\mu }=0.  \label{ConservedB}
\end{equation}
This result is sometimes referred to as `Noether's theorem'. As is well
known, `Noether's theorem' is used to connect symmetries with conserved
currents (and thence conserved charges, subject to suitable boundary
conditions).\footnote{%
For an excellent discussion of the connection between a transformation
having the status of a symmetry, in the sense of preserving the form of the
Euler-Lagrange equations, and the invariance of the action under the
transformation, see Doughty (1990, sections 9.2 and 9.5).} Confusion can
arise when we attempt to use (\ref{ConservedB}) to form conserved currents
associated with local gauge symmetries, as we will see in what follows.%
\footnote{%
Problems can also arise with respect to space-time transformations when $%
\sum_{i}\partial _{\mu }B_{i}^{\mu }=0$ is used to form a conserved current.
See, for example, Munoz (1996). See also Brading and Brown (2000) for
further discussion.}

\bigskip

\section{Global Gauge Symmetry: Noether's First Theorem}

\bigskip

In the case where the action $S$\ ($S=\int Ld^{4}x$) is invariant under a
finite dimensional continuous group of transformations depending smoothly on
$\rho $\ independent parameters $\omega _{\alpha }$, ($\alpha =1,$ 2, ... , $%
\rho $), i.e. when the symmetry is global, we can write
\begin{equation}
\delta \psi _{i}=\sum_{\alpha }\frac{\partial \left( \delta \psi _{i}\right)
}{\partial \left( \Delta \omega _{\alpha }\right) }\Delta \omega _{\alpha }
\end{equation}
where $\Delta \omega _{\alpha }$ is used to indicate that we take
infinitesimal $\omega _{\alpha }$. We substitute this into (\ref{GenExpGauge}%
), yielding
\begin{equation}
\sum_{i}\left[ \Psi \right] _{i}\frac{\partial \left( \delta \psi
_{i}\right) }{\partial \left( \Delta \omega _{\alpha }\right) }\Delta \omega
_{\alpha }\equiv -\sum_{i}\partial _{\mu }\left( \frac{\partial L}{\partial
\left( \partial _{\mu }\psi _{i}\right) }\frac{\partial \left( \delta \psi
_{i}\right) }{\partial \left( \Delta \omega _{\alpha }\right) }\Delta \omega
_{\alpha }\right) .
\end{equation}
Then, since $\Delta \omega _{\alpha }$ is not a function of space or time,
it can be removed from the derivative on the right-hand side and cancelled.
This completes the derivation of \textbf{Noether's First Theorem}, which we
now state.

\bigskip

\begin{theorem}
If the action $S$ is invariant under a finite dimensional continuous group
of transformations depending smoothly on $\rho $\ independent parameters $%
\omega _{\alpha }$, ($\alpha =1,2,...,\rho $), then there exist the $\rho $
relationships
\begin{equation}
\sum_{i}\left[ \Psi \right] _{i}\frac{\partial \left( \delta _{0}\psi
_{i}\right) }{\partial \left( \Delta \omega _{\alpha }\right) }\equiv
\partial _{\mu }j_{\alpha }^{\mu }  \label{Th1}
\end{equation}
where
\begin{equation}
j_{\alpha }^{\mu }=-\sum_{i}\frac{\partial L}{\partial \left( \partial _{\mu
}\psi _{i}\right) }\frac{\partial \left( \delta \psi _{i}\right) }{\partial
\left( \Delta \omega _{\alpha }\right) }.  \label{jmu}
\end{equation}
\end{theorem}

\bigskip

When the Euler-Lagrange field equations are assumed to be satisfied for all
the fields on which the Lagrangian depends, it follows from Noether's First
Theorem that there exist $\rho $ conserved currents, one for every parameter
on which the symmetry group depends:
\begin{equation}
\partial _{\mu }j_{\alpha }^{\mu }=0.
\end{equation}
Subject to suitable boundary conditions, this may be integrated to give a
conserved charge:
\begin{equation}
\frac{d}{dt}Q_{\alpha }=0
\end{equation}
where
\begin{equation}
Q_{\alpha }:=\int d^{3}xj_{\alpha }^{0}(x).
\end{equation}

\bigskip

Clearly, however, if $\Delta \omega _{\alpha }$ is an arbitrary function of $%
x^{\mu }$ rather than a constant, it cannot be eliminated from (\ref
{ConservedB}) in this way to give a current that is independent of the
gauge-parameter, and this is where the potential confusions begin to arise.
We consider the case of gauge symmetries depending on arbitrary functions of
$x^{\mu }$ in the following sections.

\bigskip

For a concrete example of Noether's First Theorem, consider the global gauge
symmetry of the Lagrangian associated with the Klein-Gordon equation for a
free complex scalar field:
\begin{equation}
L_{m}=\partial _{\mu }\psi \partial ^{\mu }\psi ^{\ast }-m^{2}\psi \psi
^{\ast }.  \label{Lm}
\end{equation}
$L_{m}$\ is invariant under $\psi \rightarrow \psi ^{\prime }=\psi
e^{i\theta }$, $\psi ^{\ast }\rightarrow \psi ^{\ast \prime }=\psi ^{\ast
}e^{-i\theta }$, $\theta $\ a constant, and the corresponding conserved
Noether current is
\begin{equation}
j_{L_{m}}^{\mu }=i\left( \psi ^{\ast }\partial ^{\mu }\psi -\psi \partial
^{\mu }\psi ^{\ast }\right) .
\end{equation}
This application of Noether's First Theorem to global gauge symmetry is
entirely familiar (see for example Ryder, 1996, p. 91).

\bigskip

\section{Local Gauge Symmetry: Noether's Second Theorem\label{Th2}}

\bigskip

Consider now an infinite dimensional group of transformations depending
smoothly on $\rho $ arbitrary functions $p_{\alpha }(x^{\mu })$ ($\alpha =1,$
2, ... , $\rho )$ and their first derivatives, and denote such a group by $%
G_{\infty \rho }$.\footnote{%
The restriction to the first derivative is again imposed for convenience,
and the results presented here extend straightforwardly to include higher
derivatives.} For an infinitesimal transformation of $\psi _{i}$ we can
write
\begin{equation}
\delta \psi _{i}=\sum_{\alpha }\left\{ a_{\alpha i}\left( \psi _{i},\partial
_{\mu }\psi _{i},x^{\mu }\right) \Delta p_{\alpha }(x^{\mu })+b_{\alpha
i}^{\mu }\left( \psi _{i},\partial _{\mu }\psi _{i},x^{\mu }\right) \partial
_{\mu }\left( \Delta p_{\alpha }(x^{\mu })\right) \right\}
\label{delta(phi)}
\end{equation}
where the $\Delta p_{\alpha }$ indicates that we are taking infinitesimal $%
p_{\alpha }$. We can then make use of (\ref{GenExpGauge}) to prove the
following theorem, found in Noether's 1918 paper, which we will refer to as
\textbf{Noether's Second Theorem}.\footnote{%
For details of the derivation see Noether, 1918; Trautman, 1962; Brading and
Brown, 2000.}

\bigskip

\begin{theorem}
If the action $S$\ is invariant under a group $G_{\infty \rho }$ then there
exist the $\rho $\ relationships
\begin{equation}
\sum_{i}\left[ \Psi \right] _{i}a_{\alpha i}\equiv \sum_{i}\partial _{\mu
}\left( \left[ \Psi \right] _{i}b_{\alpha i}^{\mu }\right) .
\label{2ndtheroem}
\end{equation}
\end{theorem}

\bigskip

This is derived by noticing that, since they are arbitrary, we could choose
the $p_{\alpha }$ and their derivatives so that they vanish on the boundary.
Thus, the interior contribution to VP must vanish independently of the
boundary contribution, and (\ref{2ndtheroem}) is the condition for the
vanishing of the integral associated with the interior contribution. The
theorem tells us that there are dependencies between the Lagrange
expressions $\left[ \Psi \right] _{i}$ and their derivatives. This
dependency follows from the local gauge invariance of the Lagrangian, and
its precise form depends on the particular structure of the gauge
transformation.\footnote{%
When the Lagrangian depends on a single field, the Second Theorem leads to a
contraint on the Lagrange expression. Consider, for example, classical
Maxwell electromagnetism (see Brading, 2000, for details).}

\bigskip

To make the content of this theorem concrete, consider the specific case
\begin{equation}
L=D_{\mu }\psi D^{\mu }\psi ^{\ast }-m^{2}\psi \psi ^{\ast }-\frac{1}{4}%
F^{\mu \nu }F_{\mu \nu }  \label{Ltotal}
\end{equation}
where $D_{\mu }=\partial _{\mu }+iqA_{\mu }$ is the covariant derivative,
and $F^{\mu \nu }\ $is some function of $\partial ^{\nu }A^{\mu }$ but not
of $A^{\mu }$. $L$ is invariant under local gauge transformations
\begin{equation}
\left.
\begin{array}{c}
\psi \rightarrow \psi ^{\prime }=\psi e^{iq\theta (x)} \\
\psi ^{\ast }\rightarrow \psi ^{\ast \prime }=\psi ^{\ast }e^{-iq\theta (x)}
\\
A_{\mu }\rightarrow A_{\mu }^{\prime }=A_{\mu }+\partial _{\mu }\theta (x).
\end{array}
\right\}  \label{GaugeTransf}
\end{equation}
In this case, we have only one arbitrary function $p=\theta ,$ and,
infinitesimally,
\begin{equation}
\left.
\begin{array}{c}
\delta \psi =iq(\Delta \theta )\psi \\
\delta \psi ^{\ast }=-iq(\Delta \theta )\psi ^{\ast } \\
\delta A_{\mu }=\partial _{\mu }(\Delta \theta ).
\end{array}
\right\}
\end{equation}
Hence, from (\ref{delta(phi)}) we see that

\begin{equation}
\left.
\begin{array}{c}
a_{\psi }=iq\psi \text{, }b_{\psi }^{\nu }=0 \\
a_{\psi ^{\ast }}=-iq\psi ^{\ast }\text{, }b_{\psi ^{\ast }}^{\nu }=0 \\
a_{A_{\mu }}=0\text{, }b_{A_{\mu }}^{\nu }=\delta _{\mu }^{\nu }.
\end{array}
\right\}  \label{a and b}
\end{equation}
Therefore (\ref{2ndtheroem}) yields
\begin{eqnarray}
&&\left[ \frac{\partial L}{\partial \psi }-\partial _{\mu }\left( \frac{%
\partial L}{\partial (\partial _{\mu }\psi )}\right) \right] iq\psi +\left[
\frac{\partial L}{\partial \psi ^{\ast }}-\partial _{\mu }\left( \frac{%
\partial L}{\partial (\partial _{\mu }\psi ^{\ast })}\right) \right] \left(
-iq\psi ^{\ast }\right)  \label{Th2RFT} \\
&\equiv &\partial _{\mu }\left[ \frac{\partial L}{\partial A_{\mu }}%
-\partial _{\nu }\left( \frac{\partial L}{\partial (\partial _{\nu }A_{\mu })%
}\right) \right] ,  \notag
\end{eqnarray}
from which we conclude that
\begin{equation}
\partial _{\mu }\partial _{\nu }F^{\mu \nu }\equiv 0  \label{Bianchi}
\end{equation}
where we have defined $F^{\mu \nu }$ as
\begin{equation}
F^{\mu \nu }:=\frac{\partial L}{\partial (\partial _{\nu }A_{\mu })}.
\end{equation}

Equation (\ref{Bianchi}) states that the derivative $\partial _{\nu }A_{\mu
} $ must appear in the Lagrangian in the combination $\partial _{\nu }A_{\mu
}-\partial _{\mu }A_{\nu },$ making $F^{\mu \nu }$ anti-symmetric.

\bigskip

In the few places where Noether's Second Theorem is discussed, the above
result (and its analogue in other theories) is taken to be everything that
follows from the Second Theorem. This is not the case: more can be derived
from the Second Theorem.

\bigskip

Consider first the specific example of the Lagrangian (\ref{Ltotal}). In
addition to the above result (\ref{Bianchi}), we can derive the following
from the Second Theorem by using the electromagnetic field equations
\begin{equation}
\frac{\partial L}{\partial A_{\mu }}-\partial _{\nu }\left( \frac{\partial L%
}{\partial (\partial _{\nu }A_{\mu })}\right) =0.
\end{equation}
From (\ref{Th2RFT}), we conclude that
\begin{equation}
\left[ \frac{\partial L}{\partial \psi }-\partial _{\mu }\left( \frac{%
\partial L}{\partial (\partial _{\mu }\psi )}\right) \right] \left( -iq\psi
\right) +\left[ \frac{\partial L}{\partial \psi ^{\ast }}-\partial _{\mu
}\left( \frac{\partial L}{\partial (\partial _{\mu }\psi ^{\ast })}\right) %
\right] iq\psi ^{\ast }=0
\end{equation}
and substituting in $L_{total}$\ we get
\begin{equation}
\partial _{\mu }j^{\mu }=0  \label{continuity eqn}
\end{equation}
where
\begin{equation}
j^{\mu }=iq\left( \psi ^{\ast }D^{\mu }\psi -\psi D^{\mu }\psi ^{\ast
}\right)  \label{ConservedCurrent}
\end{equation}
is the familiar electric 4-current. Hence, we see that the current
continuity equation can be derived from local gauge symmetry in conjunction
with the gauge field equations, via Noether's Second Theorem. The continuity
equation can, of course, be derived from the matter field equations, but the
Second Theorem shows that while the matter field equations are a sufficient
condition for the derivation of the continuity equation, they are not a
necessary condition (in the case of Lagrangian (\ref{Ltotal})).\footnote{%
Bergmann (1959) follows a similar procedure to that presented in this
section, without reference to Noether's second theorem. He terms the
resulting conservation laws `strong' conservation laws. These are what
Noether called `improper' conservation laws (see below in the main text).
Trautman (1962) appropriates Bergmann's term `strong' for continuity
equations that are satisfied independently of \textit{any} field equations
(see section 5).} This is in contrast to the case of global gauge symmetry,
above, where the current continuity equation associated with the Lagrangian (%
\ref{Lm}) is obtained as a consequence of the matter field equations, via
Noether's First Theorem, and where the matter field equations are \textit{%
necessary} and sufficient for deriving the continuity equation.

\bigskip

What we have here is an instance of the general result that, when the
transformations of only the gauge fields depend on $\partial _{\mu
}p_{\alpha },$ local gauge symmetry plus satisfaction of the gauge field
equations leads to a conserved current. In what follows we will see two
further methods for arriving at continuity equations such as (\ref
{continuity eqn}) via local gauge symmetry and satisfaction of the field
equations: that of Trautman (in the following section) and that of Utiyama
(see section 6).

\bigskip

We end this section with an historical aside. Noether (1918) distinguished
between `improper' and `proper' conservation laws. `Improper' conservation
laws can be derived without the field equations for the associated field
being satisfied. \ In contrast, where a necessary condition for deriving a
conservation law is that the field equations of the associated fields are
satisfied, these conservations laws are termed `proper'. This distinction is
due to Hilbert, and was made during considerations of the status of energy
conservation in General Relativity (these being what prompted Noether's 1918
work). All the results presented here for local gauge symmetry have
analogues in General Relativity, where diffeomorphism invariance is the
analogue of local gauge invariance (for further details, see Brading and
Brown, 2000). Finally, note also that when Weyl (1918) made the first
attempt to derive conservation of charge from a postulated gauge symmetry,
independently of Noether's work, his method turned out to be an instance of
Noether's Second Theorem with one set of field equations assumed to be
satisfied; he later repeated this method (Weyl, 1928 and 1929) in the
context of quantum theory (for details, see Brading, 2000).

\bigskip

\section{Global Subgroups of Local Gauge Groups\label{Rigid}}

\bigskip

In the case of a theory with local gauge symmetry where there exists a
non-trivial global subgroup, we can make use of Noether's First Theorem with
respect to this global subgroup in two ways.

\bigskip

First, we can simply apply Noether's First Theorem to global subgroups. In
the case of (\ref{Ltotal}), from application of Noether's First Theorem to
the global subgroup defined by $\theta =$ constant, with the matter field
equations assumed to be satisfied, we obtain (\ref{ConservedCurrent}) as our
conserved current once again. Restricting ourselves to the use of Noether's
First Theorem in the case of locally gauge symmetric theories is
nevertheless subtly misleading, since it suggests that satisfaction of the
matter field equations is a necessary condition for the derivation of a
conserved current. In fact, as we have seen from Noether's Second Theorem,
with respect to (\ref{Ltotal}) the conserved current is an expression of the
lack of independence of the matter and gauge fields, and can be obtained by
assuming that the gauge field equations are satisfied independently of
whether the matter field equations are satisfied.\ In the locally gauge
invariant theory, satisfaction of the matter field equations is merely a
sufficient condition for deriving the existence of a conserved current, and
not a necessary condition. This can be seen more clearly if we consider the
second way of using Noether's First Theorem with respect to a non-trivial
global subgroup of a local symmetry group.

\bigskip

Trautman (1962) combines Noether's First Theorem with Noether's Second
Theorem in the case where there exists a non-trivial global subgroup defined
by
\begin{equation}
\Delta p_{a}(x^{\mu })=\Delta \omega _{a}.
\end{equation}
Then (\ref{delta(phi)}) becomes
\begin{equation}
\delta \psi _{i}=a_{\alpha i}\Delta \omega _{a}+b_{\alpha i}^{\mu }\partial
_{\mu }\left( \Delta \omega _{a}\right) =a_{\alpha i}\Delta \omega _{a}
\end{equation}
since $\partial _{\mu }\left( \Delta \omega _{a}\right) =0.$ Substituting
this into the first theorem (\ref{Th1}), we get:
\begin{equation}
\sum_{i}\left[ \Psi \right] _{i}a_{\alpha i}\equiv \partial _{\mu }j_{\alpha
}^{\mu }.
\end{equation}
But from the second theorem (\ref{2ndtheroem})
\begin{equation*}
\sum_{i}\left[ \Psi \right] _{i}a_{\alpha i}\equiv \sum_{i}\partial _{\mu
}\left( \left[ \Psi \right] _{i}b_{\alpha i}^{\mu }\right)
\end{equation*}
and hence
\begin{equation}
\partial _{\mu }j_{\alpha }^{\mu }\equiv \sum_{i}\partial _{\mu }\left(
\left[ \Psi \right] _{i}b_{\alpha i}^{\mu }\right)
\end{equation}
therefore
\begin{equation}
\partial _{\mu }\left\{ j_{\alpha }^{\mu }-\sum_{i}\left( \left[ \Psi \right]
_{i}b_{\alpha i}^{\mu }\right) \right\} \equiv 0.  \label{TrautmanEqn}
\end{equation}
Trautman calls such expressions `strong' conservation laws because they are
derived independently of any equations of motion.\footnote{%
Trautman takes the term from Bergmann (1949) but, as noted above in a
footnote, Bergmann applies the term `strong' to conservation laws for which
the field equations \textit{of the fields associated with the conservation
law} are not necessary for the conservation law, even though other field
equations are necessarily assumed to be satisfied as part of the derivation.}

\bigskip

In the case of the Lagrangian (\ref{Ltotal}), (\ref{TrautmanEqn}) yields
\begin{equation}
0\equiv \partial _{\mu }\left\{ j^{\mu }-\partial _{\nu }\left[ \frac{%
\partial L}{\partial A_{\mu }}-\partial _{\nu }\left( \frac{\partial L}{%
\partial \left( \partial _{\nu }A_{\mu }\right) }\right) \right] \delta
_{\mu }^{\nu }\right\} .  \label{Trautman2}
\end{equation}
Hence,
\begin{equation}
\partial _{\mu }\partial _{\nu }F^{\mu \nu }\equiv 0.
\end{equation}
Or, returning to (\ref{Trautman2}), if the gauge field equations are
satisfied, we have the conclusion that
\begin{equation*}
\partial _{\mu }j^{\mu }=0.
\end{equation*}
These results are obtainable by substituting (\ref{Ltotal}) directly into
the Second Theorem, as we have seen. We mention Trautman's result here in
part because it makes vivid the point that, in the case of the locally gauge
invariant Lagrangian (\ref{Ltotal}), the matter field equations are not a
necessary condition for deriving the existence of a conserved matter-field
current $j^{\mu }$: the continuity equation can be derived via (\ref
{Trautman2}) by assuming that the gauge field equations are satisfied.

\bigskip

We will see in the following section that (\ref{TrautmanEqn}) is derivable
without the assumption that there exists a non-trivial global subgroup.

\bigskip

\section{Local Gauge Symmetry: Theorem 3\label{FieldEqns}}

\bigskip

In this section, we present results due to Utiyama in the form of a theorem;
we set this in the context of the results already described in this paper,
and draw attention to a corollary due to Al-Kuwari and Taha that we consider
to be of particular interest, namely the derivation of the `Coupled Field
Equations' in their general form.

\bigskip

We have seen that we cannot follow the procedure used for global gauge
symmetries in the case of local gauge symmetries to form gauge-independent
currents. A current that is dependent on $\Delta p_{\alpha }$ is not
satisfactory - in particular, such gauge-dependent quantities are not
observable.\footnote{%
How to deal with this problem is the main subject of the exchange cited in
the Introduction, between Karatas and Kowalski (1990) and Al-Kuwari and Taha
(1991). Not everyone agrees that gauge-dependent quantities are problematic,
however; see for example Bak, Cangemi and Jackiw (1994).} A question that
Noether did not address is whether useful, gauge-independent results can be
derived from considering the boundary contribution to VP in the case of
local symmetries. Al-Kuwari and Taha (1991) consider just this problem,
drawing heavily on the work of Utiyama (1956), and citing Frampton's (1987)
discussion of Utiyama. No reference to Noether's Second Theorem is made in
any of these cases. Utiyama (1956, 1959) starts by retaining both the
interior and boundary contributions to VP, and here we follow this more
general approach. The Al-Kuwari and Taha results arise when we add the
assumption that the Euler-Lagrange equations are satisfied for all the
fields on which the Lagrangian depends, as will be indicated.

\bigskip

\begin{theorem}
If the action $S$\ is invariant under an infinite dimensional continuous Lie
group depending smoothly on $\rho $\ arbitrary functions $p_{\alpha }(x^{\mu
})$\ ($\alpha =1,\ 2,...,\rho )$\ and their first derivatives, then there
exist three sets of $\rho $ relationships:
\begin{equation}
\sum_{i}\left[ \Psi \right] _{i}a_{\alpha i}\equiv -\sum_{i}\partial _{\mu
}\left( \frac{\partial L}{\partial \left( \partial _{\mu }\psi _{i}\right) }%
a_{\alpha i}\right)   \label{1}
\end{equation}
\begin{equation}
\sum_{i}\left[ \Psi \right] _{i}b_{\alpha i}^{\mu }\equiv -\sum_{i}\left[
\frac{\partial L}{\partial \left( \partial _{\mu }\psi _{i}\right) }%
a_{\alpha i}+\partial _{\mu }\left( \frac{\partial L}{\partial \left(
\partial _{\mu }\psi _{i}\right) }b_{\alpha i}^{\mu }\right) \right]
\label{2}
\end{equation}
\begin{equation}
0\equiv \sum_{i}\left[ \frac{\partial L}{\partial \left( \partial _{\mu
}\psi _{i}\right) }b_{\alpha i}^{\mu }+\frac{\partial L}{\partial \left(
\partial _{\nu }\psi _{i}\right) }b_{\alpha i}^{\nu }\right] .  \label{3}
\end{equation}
\end{theorem}

\bigskip

Notice that in the special case $p_{\alpha }(x^{\mu })=\omega _{\alpha },$ (%
\ref{1}) reduces to (\ref{Th1}), and we recover Noether's First Theorem.

\bigskip

The derivation of Theorem 3 proceeds as follows. We begin from the general
expression (\ref{GenExpGauge}) given in section 2 (the common starting point
of Noether's two theorems), and we substitute (\ref{delta(phi)}) into this
via (\ref{B(fields only)}), yielding the expressions
\begin{equation}
\sum_{i}\left[ \Psi \right] _{i}\left( a_{\alpha i}\Delta p_{\alpha
}+b_{\alpha i}^{\mu }\partial _{\mu }\left( \Delta p_{\alpha }\right)
\right) \equiv -\sum_{i}\partial _{\mu }\left\{ \frac{\partial L}{\partial
\left( \partial _{\mu }\psi _{i}\right) }\left( a_{\alpha i}\Delta p_{\alpha
}+b_{\alpha i}^{\mu }\partial _{\mu }\left( \Delta p_{\alpha }\right)
\right) \right\} ,
\end{equation}
one for every $\rho $ independent arbitrary functions on which the symmetry
group depends. For each of these $\rho $ expressions, we proceed by
collecting terms in $\Delta p_{\alpha }$ and its derivatives:
\begin{eqnarray}
&&\sum_{i}\left[ \Psi \right] _{i}a_{\alpha i}\Delta p_{\alpha }+\sum_{i}%
\left[ \Psi \right] _{i}b_{\alpha i}^{\mu }\partial _{\mu }\left( \Delta
p_{\alpha }\right) \\
&\equiv &-\sum_{i}\partial _{\mu }\left( \frac{\partial L}{\partial \left(
\partial _{\mu }\psi _{i}\right) }a_{\alpha i}\right) \Delta p_{\alpha }
\notag \\
&&-\sum_{i}\left[ \frac{\partial L}{\partial \left( \partial _{\mu }\psi
_{i}\right) }a_{\alpha i}+\partial _{\mu }\left( \frac{\partial L}{\partial
\left( \partial _{\mu }\psi _{i}\right) }b_{\alpha i}^{\mu }\right) \right]
\partial _{\mu }\left( \Delta p_{\alpha }\right)  \notag \\
&&-\sum_{i}\left[ \frac{\partial L}{\partial \left( \partial _{\mu }\psi
_{i}\right) }b_{\alpha i}^{\mu }+\frac{\partial L}{\partial \left( \partial
_{\nu }\psi _{i}\right) }b_{\alpha i}^{\nu }\right] \partial _{\nu }\partial
_{\mu }\left( \Delta p_{\alpha }\right) .  \notag
\end{eqnarray}
But $\Delta p_{\alpha }$ and its derivatives are arbitrary, and hence the
coefficients of $\Delta p_{\alpha }$ and its derivatives must vanish
independently, enabling us to extract three separate equations and formulate
Theorem 3.

\bigskip

Comparing equations (\ref{1}), (\ref{2}) and (\ref{3}) with those of Utiyama
(1959, p. 24), we see that his second and third results are simply (\ref{2})
and (\ref{3}), but his first result is different. Utiyama's (1959, p. 24)
results are obtained by observing that the interior and boundary
contributions must vanish independently,\footnote{%
This is because the functions $p_{\alpha }$ are arbitrary, and so we could
choose that the $p_{\alpha }$ and their derivatives vanish on the boundary.
Therefore, the interior contribution must vanish independently of what
happens on the boundary, and since the entire variation must vanish in all
cases, the boundary contribution must vanish even when the arbitrary
functions are not chosen to vanish on the boundary.} and by focusing on the
boundary contribution. Noether's Second Theorem is the condition for the
vanishing of the interior, and we can substitute (\ref{2ndtheroem}) into (%
\ref{1}) to obtain Utiyama's first result (1959, p. 24, equation 2.6):
\begin{equation}
\sum_{i}\partial _{\mu }\left( \frac{\partial L}{\partial \left( \partial
_{\mu }\psi _{i}\right) }a_{\alpha i}+\left[ \Psi \right] _{i}b_{\alpha
i}^{\mu }\right) \equiv 0.  \label{Utiyama1}
\end{equation}

\bigskip

The significance of the results (\ref{1}) and (\ref{Utiyama1}) can be
understood as follows. Consider the special case where the fields $\psi _{i}$
on which the Lagrangian depends divide into two sets: one whose gauge
transformations depend on\ $p_{\alpha }$ but not on $\partial _{\mu
}p_{\alpha },$ the other whose transformations depend on $\partial _{\mu
}p_{\alpha }$\ but not on $p_{\alpha }.$ Then it follows from (\ref{1}) and (%
\ref{Utiyama1}) that if either set of field equations is satisfied, there
exists a conserved current of the form$\ $%
\begin{equation}
j_{\alpha }^{\mu }:=-\sum_{i}\frac{\partial L}{\partial \left( \partial
_{\mu }\psi _{i}\right) }a_{\alpha i}.  \label{General Jmu}
\end{equation}
The Lagrangian (\ref{Ltotal}) is just such a special case: as discussed in
section 4, above, the continuity equation for the electric current can be
derived from satisfaction of either the matter field equations or the gauge
field equations.\bigskip

Equation (\ref{Utiyama1}) is the result obtained by Trautman (when we
substitute in (\ref{General Jmu})) via consideration of the global subgroup
of the local gauge group $p_{\alpha }(x^{\mu })=\omega _{\alpha }$ (see
section 5, above): we see here that Trautman's result is derivable more
generally.

\bigskip

Turning now to equation (\ref{2}), we first note Utiyama's (1959, p. 27,
equation 2.14) observation that in the specific case of the Lagrangian (\ref
{Ltotal}) we obtain:
\begin{equation}
\sum_{i}\frac{\partial L}{\partial \left( \partial _{\mu }\psi _{i}\right) }%
a_{\alpha i}\equiv \frac{\partial L}{\partial A_{\mu }}.
\label{current identity}
\end{equation}
Thus, the current associated with the matter fields equations (on the
left-hand side of (\ref{current identity})) is identified with the current
of Maxwell's equations with sources (on the right-hand side of (\ref{current
identity})). In more general terms, when condition (\ref{current identity})
is satisfied the matter-field current associated with the Lagrangian acts as
the source for the gauge fields.

\bigskip

There is, however a wider and completely general significance to (\ref{2}),
which we turn to in the following section.

\bigskip

The significance of the final set of equations presented in Theorem 3, (\ref
{3}), is seen most clearly from the specific example of the Lagrangian (\ref
{Ltotal}). In this case (see Utiyama, 1959, p. 27), (\ref{3}) says that the
derivative $\partial _{\nu }A_{\mu }$ must appear in the Lagrangian in the
combination $\partial _{\nu }A_{\mu }-\partial _{\mu }A_{\nu },$ or in other
words (\ref{Bianchi}). More generally, we have a restriction on those fields
whose transformations depend on $\partial _{\mu }p_{\alpha }.$

\bigskip

The three sets of equations given by Al-Kuwari and Taha (1991, equations
34), modulo some technical details, are arrived at from (\ref{1}), (\ref{2})
and (\ref{3}) by assuming that the Euler-Lagrange equations are satisfied
for all the fields on which the Lagrangian depends, i.e. $\sum_{i}\left[
\Psi \right] _{i}=0$, so that the left-hand sides of (\ref{1}) and (\ref{2})
vanish.\footnote{%
At this point, it is perhaps worth offering the following cautionary remark.
Throughout this paper we have been careful to write that a sufficient
condition for $\sum_{i}\left[ \Psi \right] _{i}$ vanishing is that the
Euler-Lagrange equations are satisfied \textit{for all the fields on which
the Lagrangian depends}. More precisely still, we restrict our attention to
those fields that feature in the symmetry transformation under
consideration. In the presence of `background fields' or `absolute objects'
that participate in the symmetry transformation, satisfaction of the
Euler-Lagrange equations may not be sufficient for the vanishing of all the
Lagrange expressions $\left[ \Psi \right] _{i}$, and hence $\sum_{i}\left[
\Psi \right] _{i}$ may not be zero. For example, consider the locally gauge
invariant Lagrangian
\begin{equation*}
L_{1}(\psi ,\partial _{\mu }\psi ,\psi ^{\ast },\partial _{\mu }\psi ^{\ast
},A_{\mu },x^{\mu })=D_{\mu }\psi D^{\mu }\psi ^{\ast }-m^{2}\psi \psi
^{\ast }.
\end{equation*}
In this case, $\left[ \Psi \right] _{A_{\mu }}=\frac{\partial L_{1}}{%
\partial A_{\mu }}-0=j^{\mu },$ and the left-hand side of equation (\ref{2})
does not vanish even when all the Euler-Lagrange equations associated with $%
L_{1}$ are satisfied. Failure to notice this would lead to inconsistent
results.} In our opinion, their second set of equations is their most
important result. We discuss this in detail in the following section, but
first some brief remarks about the other two sets of equations. Their first
set of equations is potentially misleading with respect to the necessary and
sufficient conditions for deriving the existence of the conserved current:
their derivation assumes satisfaction of \textit{both} the matter field
equations and the gauge field equations, but satisfaction of \textit{either}
set of field equations is sufficient for the conserved current to be
derived, as we saw from Noether's Second Theorem in section 4 above.
Nevertheless, as Al-Kuwari and Taha emphasise, the first set of equations
does make clear the important point that no further conserved currents can
be derived from local gauge symmetry than from global gauge symmetry when
the Euler-Lagrange equations are assumed to be satisfied. The third set of
equations is also potentially misleading: Al-Kuwari and Taha assume from the
outset that the Euler-Lagrange equations are satisfied, so their version of (%
\ref{3}) appears to depend on the satisfaction of these equations; recall,
however, that (\ref{3}) can be derived from the local gauge invariance of
the Lagrangian independently of any Euler-Lagrange equations, as we also saw
from Noether's Second Theorem in section 4 above.

\bigskip

\subsection{Coupled Field Equations}

\bigskip

To bring out the wider significance of (\ref{2}) we turn to Al-Kuwari and
Taha.\footnote{%
Although consistent with the spirit of Al-Kuwari and Taha's results, the
discussion presented here differs from theirs in various technical respects.}
When the Euler-Lagrange equations are assumed to be satisfied for all the
fields on which the Lagrangian depends (or, more precisely, for all the
fields whose transformations depend on $\partial _{\mu }p_{\alpha })$,
equation (\ref{2}) of Theorem 3 yields what we may call the \textbf{`Coupled
Field Equations'}:
\begin{equation}
j_{\alpha }^{\mu }=\sum_{i}\partial _{\sigma }\left( F^{i\mu }b_{\alpha
i}^{\sigma }\right)  \label{CFE}
\end{equation}
where
\begin{equation}
j_{\alpha }^{\mu }:=-\sum_{i}\frac{\partial L}{\partial \left( \partial
_{\mu }\psi _{i}\right) }a_{\alpha i}
\end{equation}
and
\begin{equation}
F^{i\mu }:=\frac{\partial L}{\partial \left( \partial _{\mu }\psi
_{i}\right) }.
\end{equation}
We term (\ref{CFE}) `Coupled Field Equations' because of the form of the
inter-relationship they describe between the different fields appearing in
the Lagrangian. In the specific example we have been considering, the
Lagrangian (\ref{Ltotal}), (\ref{CFE}) becomes
\begin{equation}
j^{\mu }=\partial _{\nu }F^{\nu \mu }  \label{Maxwell}
\end{equation}
where
\begin{equation}
F^{\nu \mu }=\frac{\partial L}{\partial \left( \partial _{\mu }A_{\nu
}\right) }.
\end{equation}
Condition (\ref{Maxwell}) tells us that the vanishing of the boundary
contribution to the variation in the action requires a balance between the
current associated with the matter fields and the propagation of the gauge
fields$.$

\bigskip

Notice the important point that the general form of these coupled field
equations, (\ref{CFE}), has been derived independently of the form of any
specific Lagrangian or Euler-Lagrange equations$.$ We simply assume that the
Lagrangian is invariant under local gauge transformations of the general
form (\ref{delta(phi)}), and that the field equations are satisfied, but we
don't have to know what the field equations are in order to derive the
general form of the coupled field equations.

\bigskip

This concludes the results derivable from Noether's variational problem for
global and local gauge symmetries.\footnote{%
The results are straightforwardly generalisable to the case where $\delta
x\neq 0$ (see Doughty, 1990; Utiyama, 1959; and Brading and Brown, 2000).}

\bigskip

\bigskip

{\Large Acknowledgements}

\bigskip

We would like to thank Roman Jackiw and Antigone Nounou for discussion of
some aspects of this paper. One of us (K.B.) thanks the A. H. R. B. and St.
Hugh's College, Oxford, for financial support.

\bigskip

{\Large References}

\bigskip

Al-Kuwari, H. A., and Taha, M. O., 1991: `Noether's theorem and local gauge
invariance', \textit{American Journal of Physics}, \textbf{59} (4), 363-365.

Bak, D., Cangemi. D., and Jackiw, R., 1994: `Energy-momentum conservation in
gravity theories', \textit{Physical Review D}, \textbf{49} (10), 5173-81.

Bergmann, P. G., 1949: `Non-linear Field Theories', \textit{Physical Review}
\textbf{75} (4), 680-685.

Brading, K. A., 2000: `Which Symmetry? Noether, Weyl, and Conservation of
Electric Charge', \textit{Studies in History and Philosophy of Modern Physics%
}, forthcoming.

Brading, K. A., and Brown, H. R., 2000: `Noether's\ Theorems', in
preparation.

Doughty, N. A., 1990: \textit{Lagrangian Interaction}, Addison-Wesley
Publishers Ltd.

Frampton, P. H., 1987: \textit{Gauge Field Theories}, Benjamin/Cummings.

Karatas, D. L., and Kowalski, K. L., 1990: `Noether's theorem and local
gauge transformations', \textit{American Journal of Physics}, \textbf{58}
(2), 123-131.

Munoz, G., 1996: `Lagrangian field theories and energy-momentum tensors',
\textit{American Journal of Physics}, \textbf{64} (9), 1153-1157.

Noether, E., 1918: `Invariante Variationsprobleme', \textit{Nachr. d. Konig.
Gesellsch. d. Wiss. zu Gottingen, Math-phys. Klasse}, 235-257. English
translation: Tavel, 1971.

O'Raifeartaigh, L., 1997: \textit{The Dawning of Gauge Theory}, Princeton
University Press.

Ryder, L. H., 1996: \textit{Quantum Field Theory}, Second Edition, Cambridge
University Press.

Tavel, M. A., 1971: 'Noether's theorem', \textit{Transport Theory and\
Statistical Physics} \textbf{1}(3), 183-207.

Trautman, A., 1962: `Conservation Laws in General Relativity', in \textit{%
Gravitation: An Introduction to Current Research}, ed. L. Witten, John Wiley
and Sons.

Utiyama, R., 1956: `Invariant Theoretical Interpretation of Interaction',
\textit{Physical Review}, \textbf{101} (5), 1597-1607.

Utiyama, R., 1959: `Theory of Invariant Variation and the Generalized
Canonical Dynamics', \textit{Progr. Theor. Phys. Suppl.} \textbf{9}, 19-44.

Weyl, H., 1918: `Gravitation and Electricity'; English translation available
in O'Raifeartaigh,1997.

Weyl, H., 1928: \textit{The Theory of Groups and Quantum Mechanics}, English
translation, Dover edition 1950.

Weyl, H., 1929: `Electron and Gravitation'; English translation available in
O'Raifeartaigh, 1997.

\bigskip

\end{document}